\documentclass{article}

\usepackage[pdftex]{graphicx}
\usepackage[cmex10]{amsmath}
\usepackage[caption=false,font=footnotesize]{subfig}
\usepackage{fixltx2e}
\usepackage{url}
\usepackage[format=plain,textformat=simple,justification=justified,small,parskip=0cm]{caption}
\usepackage[left=3cm,top=2cm,right=2cm,bottom=2cm,nohead]{geometry}

\newcommand\erfc{\mathrm{erfc}}

\begin{document}
\title{Agent-based Versus Macroscopic Modeling of Competition and Business Processes in Economics and Finance}

\author{Aleksejus Kononovicius, Vygintas Gontis, Valentas Daniunas}

\maketitle

\begin{abstract}
We present examples of agent-based and stochastic models
of competition and business processes in economics and finance. We start from
as simple as possible models, which have microscopic, agent-based,
versions and macroscopic treatment in behavior. Microscopic and
macroscopic versions of herding model proposed by Kirman and Bass
new product diffusion are considered in this contribution as
two basic ideas. Further we demonstrate that general herding behavior
can be considered as a background of nonlinear stochastic model of financial
fluctuations.
\end{abstract}

\section{Introduction}
Statistically reasonable models of social and economic systems, first of all
stochastic and agent-based, are of great interest for a wide scientific community of interdisciplinary researchers dealing with diversity
of complex systems \cite{Daniunas2011ICCGI, Helbing2008Springer, Pietronero2008EPN, Waldrop1992SAS}.
Computer modeling is one of the key aspects of modern science, be it
physical or social or economic science, \cite{Ince2012Nature, Niemeyer2012Ars}.
In case of complex system modeling it serves as a technique in the quest for
the understanding of
the interrelation between microscopic interactions of individual agents and macroscopic,
colective, dynamics of the whole complex system. Nevertheless, some general theories or
methods that are well developed in the natural and physical sciences can be helpful
in the development of consistent micro and macro modeling of
complex systems
\cite{Pietronero2008EPN, Waldrop1992SAS, Axelrod1998Comp, Helbing2010SciCu, Johnson2010EJSSR}.

As computer modeling is very prominent and important in modern science, we start
this paper by discussing our online publishing and collaboration platform, see Section \ref{sec:web}.
The open-source applets made available online on the website ``Physics of Risk'', see \cite{RizikosFizika},
allow to reproduce most of the results presented in this paper. This is very important as reproducibility
of the results is one of the key demands in scientific society \cite{Ince2012Nature, Niemeyer2012Ars}.

From the Section \ref{sec:modelsStart} we start discussing various models applicable in economics and finance,
which highlight the important correspondence between microscopic, agent-based, and macroscopic, stochastic,
modeling. In the opening Section \ref{sec:kirman},
we present Kirman's agent-based model (see \cite{Kirman1993QJE} for original paper) and
derive its stochastic alternative, which was also done by Alfarano et al. in \cite{Alfarano2005CompEco}
using a more complex manner. In the Section \ref{sec:bass} we show that modified, unidirectional,
Kirman's agent based model can be seen as microscopic alternative to the widely known Bass
diffusion model \cite{Bass1969ManSci}. Further, in Section \ref{sec:financialmark}, we apply the
stochastic treatment of the Kirman's model for financial markets and obtain stochastic model of
absolute return similar to the CEV process \cite{Jeanblanc2009Springer} and earlier proposed model
of $1/f$ noise \cite{Kaulakys2009JStatMech, Gontis2010Sciyo, Ruseckas2010PhysRevE}. In the
Section \ref{sec:mfdfa} we show that Kirman's model possesses multifractal features, which are
seen as an important feature of many natural phenomena \cite{Kantelhardt2002PhysA}. Section \ref{sec:bursts}
closes presented discussion with some definitions and results regarding burst duration statistics generated by the class of nonlinear SDE and observable in the financial markets.

In the last section, Section \ref{sec:conclusions}, we sum up everything discussed
in this paper and share some ideas on future developments of the discussed research.

\section{Review of the related works}

Current on-going financial economic crisis provoked many papers calling for a revolution
of economical thought and emphasizing a need for a wider applications of statistical physics in the research of social complexity
\cite{Pietronero2008EPN, Helbing2010SciCu, Johnson2010EJSSR, Bouchaud2008Nature,
Bouchaud2009PhysWorld, Bouchaud2010Siena, Farmer2009Nature, Lux2009NaturePhys, Newman2011AJP,
Schinckus2010AJP}. Most of them pointing out that agent based models are very important if one
wants to effectively understand what is going on in the complex social and economic systems and
the physical intuition might provide the important bridging between the macroscopic and microscopic
modeling. These ideas somewhat traceback to the thoughts put down by Waldrop and Axelrod in
the 1990s (see \cite{Waldrop1992SAS, Axelrod1998Comp}).

In the recent decades there were many attempts to create an agent-based model for
the financial markets, yet no model so far is realistic enough and tractable to be considered as an ideal model
\cite{Cristelli2010Fermi}. One of the best examples of realistic models is so-called
Lux and Marchesi model \cite{Lux1999Nature}, which is heavily based
on the behavioral economics ideas mathematically put down as utility functions for the agents in the
market, thus it is considered to be very reasonable and realistic \cite{Cristelli2010Fermi}. Yet this model is
too complex, namely it has many
parameters and complex agent interaction mechanics, to be analytically tractable. Another example
of a very complex agent-based model would be Bornholdt's spin model \cite{Bornholdt2001IJMPC,
Kaizoji2002PhysA}, which is based on a certain interpretation of the well-known Ising model
(for the details on the original model see any handbook on statistical physics
(ex., \cite{Sethna2009Clarendon})).

Some might argue that agent-based models need not to be analytically tractable
and in fact that agent-based models are best suited to model phenomena, which is too complex
to be analytically described \cite{Bonabeau2002ProcNAS}. But the recent developments
show that many groups attempt to build a bridge between microscopic and macroscopic models.
Possibly one of the earliest attempts to do so started from not so realistic, nor tractable ``El Farol
bar problem'' \cite{Arthur1994AER}. This simple model quickly became known as the Minority
Game \cite{Challet1997PhysA} and over few years received analytic treatment
\cite{Challet2000PhysRevLett}. Another prominent simple agent-based model was created by Kirman
\cite{Kirman1993QJE}, which gained broader attention only very recently
\cite{Alfarano2005CompEco,Alfarano2008JEcoDyC,Alfi2009EPJB1,Alfi2009EPJB2}. In \cite{Kononovicius2012PhysA}
we have given this model and extended analytical treatment and have shown that this model coincides
with some prominent macroscopic, namely stochastic, models of the financial markets (see
Section \ref{sec:kirmanfinance} of this work for more details). Another interesting development was made
by following the aforementioned Bornholdt's spin model, which has recently received an analytical
treatment via mean-field formalism \cite{Krause2012PhysRevE}.

Our work in the modeling of complex social and economic systems has begun from the applications of nonlinear
stochastic differential equations (abbr. SDE) seeking reproduce statistics of
financial market data. The proposed class of equations has power law statistics
evidently very similar to the ones observed in the empirical data. As all of
this work (for broad review see \cite{Gontis2010Sciyo}) was done by relying on the
macroscopic phenomenological reasoning,
we are now motivated to find the microscopic reasoning for the proposed equations.
The development of the macroscopic treatments for the well
established agent-based models appears to be the most consistent
approach, as the movement in the opposite direction
seems to be very complex and ambiguitious task. Thus we decided that we should select the simple
agent-based models, which would have an expected macroscopic
description. In this contribution we present a few examples of the
agent-based modeling, based on the Kirman's model, in the business and finance while showing that
the examples have useful and informative macroscopic treatments.

Kirman's ant colony model \cite{Kirman1993QJE} is an agent-based
model used to explain the importance of herding inside the ant colonies
and economic systems (see the later works by Kirman (ex. \cite{Kirman2002SNDE}) and other
authors, which develop on this idea, \cite{Alfarano2005CompEco, Alfarano2008JEcoDyC}).
The analogy can be drawn as human crowd behavior is ideologically and
statistically similar in many senses. On our website, \cite{RizikosFizika},
we have presented interactive realizations of the original
Kirman's agent-based model (see \cite{RizikosFizikaKirmanAnts}),
of its stochastic treatment by Alfarano et al. \cite{Alfarano2005CompEco}
(see \cite{RizikosFizikaKirmanAntsStochastic}) and of its treatment in the
financial market scenario done by our group \cite{Kononovicius2012PhysA}
(see \cite{RizikosFizikaHerdModelFinMark, RizikosFizikaMultiFrac}).

The diffusion of new products is one of the key problems in marketing research,
and also one of the fields where we see that Kirman's model might be applied.
The Bass diffusion model is a very prominent model related
to this problem. This model is formulated as an ordinary differential equation,
which might be used to forecast the number of adopters of the new
successful product or service \cite{Bass1969ManSci}. There were suggestions
that such basic macroscopic description in marketing research can be studied
using the agent-based modeling as well \cite{Mahajan1993NorthHolland}. Thus it
is a great opportunity to explore the correspondence between the micro
and macro descriptions looking for the conditions under which both
approaches converge. The Bass Diffusion model is of great interest for
us as representing very practical and widely accepted area of
business modeling. Web based interactive models, presented on the
site \cite{RizikosFizikaBusiness} serve as an additional research
instrument available for very wide community. On our website we also
provide an interactive applet for the Bass diffusion treatment in terms
of the modified Kirman's model \cite{RizikosFizikaKirmanUniDir} (for details
on modification see Section \ref{sec:bass} of this work).

Another interesting problem tackled in this work is related to the dynamics of
the intermittent behavior. This kind of behavior is observed in many different complex
systems ranging from the geology (ex., earthquakes \cite{Corral2004PhysRevLett})
and astronomy (ex., sunspots \cite{Wheatland1998AstrophysJ}) to
the biology (ex., neuron activity \cite{Kemuriyama2010BioSys}) and finance
\cite{Cont2001RQUF}. Great review of the universality of the bursty behavior is
given by Karsai et al. \cite{Karsai2012NIH} and Kleinberg \cite{Kleinberg2003DataMin}.
In  \cite{Karsai2012NIH} the bursting behavior is considered as a point process with threshold mechanism.
In this contribution we analyze the class of nonlinear SDE exhibiting power law statistics and bursting behavior, 
which was derived from the multiplicative point process \cite{Gontis2004PhysA343,
Kaulakys2005PhysRevE,Ruseckas2011EPL} with applications for the modeling of trading activity in financial markets \cite{Gontis2006JStatMech,Gontis2008PhysA}.
This provides a very general, via
hitting time formalism \cite{Jeanblanc2009Springer,Gardiner1997Springer,Redner2001Cambridge}, approach to the modeling of bursty behavior of trading activity and absolute return in the financial markets \cite{Gontis2012ACS}.

\section{Web platform}
\label{sec:web}

Our web site \cite{RizikosFizika} was setup using WordPress
webloging software \cite{wordpress}. The setup pays to be user-friendly, powerful
and easily extensible web publishing platform, which with some effort
can be adapted to the scientist's needs. There is a wide choice of
plugins, which enable convenient usage of equations (mostly using
LaTeX). While during the setup we found that bibliography
management plugins were lacking at the time.

To accommodate our needs for equations we have worked on
improving WP-Latex plugin (available from \cite{wplatex}). Namely
we have introduced a possibility to write equations in both inline and
ordinary math modes. Implemented equation labeling, numbering
and referencing. And finally fixed some noticeable problems with vertical
placement of the inline mode equations.

Another important task was to implement bibliography management
and citations. For this cause we have used the bibtexParse PHP code
(available from \cite{bibtexParse}) to setup BiBTeX backend. From
this point on we have written our own original PHP code to link
between bibtexParse, our database and WordPress. By using this
plugin we can now easily manage and present our own papers (ex. generate
our own bibliographies), papers we have read (tag them with keywords, write
our own comments and etc.) and also communicate with the visitors
using numerous citations.

Interactive models themselves are independent from the publishing
framework. Most of them were implemented using the Java
applet technology. Some of the applets were created using
multi-paradigm simulation software AnyLogic \cite{Anylogic}, while
the others were programmed from scratch using Java programing language
\cite{Java}. AnyLogic was used in the most of agent-based scenarios
as it is a very convenient tool for agent-based modeling, while
programing from scratch gave us more control over the applets behavior
needed while doing stochastic modeling.

Either way by compiling appropriate files one obtains Java applets,
which can be included in to the articles written using WordPress. This
way articles become interactive - visitor can both theoretically
familiarize himself with the model and test if the claims made in
the post describing model were true. This happens in the same
browser window, thus the transition between theory and modeling
appears to be seamless. Due to the fact that models are implemented
as Java applets all of the numerical evaluation occurs on client machine,
while the visitor must have Java Runtime Environment installed, and
server load stays minimal. The requirement for JRE might appear to
be cumbersome, but the technology is somewhat popular and freely
available from Oracle Corp.

One of the goals of developing these models on the web site
was to provide theoretical background
of Bass Diffusion model and discuss practical steps on how such
computer simulations can be created even with limited IT knowledge
and further applied for varying purposes (see \cite{RizikosFizikaBusiness}). Thus, we have targeted
small and medium enterprises to encourage them to use modern
computer simulation tools for business planning, sale forecasting
and other purposes.

Consequently computer models and their corresponding descriptions
published at the \cite{RizikosFizikaBusiness} provide a relatively easy
starting point to get acquainted with computer simulation in business.
The published content enables site visitors to familiarize themselves with
these models interactively, running the applets directly in a browser
window, changing the parameter values and observing results. This
significantly increases accessibility and dissemination of these
simulations.

Our web site also offers another level of reproducibility by
including source code files inside the Java applet files. In this way any
willing user may use modern archiver software (ex., 7Zip) to obtain
the source code. After doing so one can analyze source code and more
deeply understand the presented models and their implementations.
This is a very important level of reproducibility in the modern scientific
context \cite{Ince2012Nature, Niemeyer2012Ars}.

\section{Extended macroscopic treatment of Kirman's model}
\label{sec:modelsStart}
\label{sec:kirman}

There is an interesting phenomenon concerning behavior of ant
colony. It appears that if there are two identical food sources
nearby, or two identical paths to the same food source
(the experiment done by Pasteels and Deneubourg
\cite{Pastels1987Birkhauser1,Pastels1987Birkhauser2}),
ants exploit only one of them at a given time. Evidently the
food source which will be used at a given time is not
certain. It is so as switches between
food sources occur, though the food sources, or paths, remain the
same.

One could assume that those different food sources are
different trading strategies or, if putting it simply, the actions
available to traders. Thus, one could argue that speculative bubbles
and crashes in the financial markets are of similar nature as the
exploitation of the food sources in ant colonies - as quality
of stock and quality of food in the ideal case can be assumed to
be constant. Thus, model \cite{Kirman1993QJE} was created using ideas
obtained from the ecological experiments
\cite{Pastels1987Birkhauser1,Pastels1987Birkhauser2}
can be applied towards the financial market modeling.

Kirman, as an economist, actually developed this model as
a general framework in context of economic modeling (see
\cite{Kirman1993QJE,Kirman2002SNDE} and his other works).
Recently his framework was also used by other authors who are concerned
with the financial market modeling (see
\cite{Alfarano2005CompEco,Alfarano2008JEcoDyC}).
Thus basing ourselves on the main ideas of these
authors and our previous results in stochastic modeling (see
\cite{Gontis2010Sciyo}) we introduce specific modifications of
Kirman's model providing a class of nonlinear stochastic
differential equations \cite{Ruseckas2010PhysRevE} applicable for the
financial variables.

Kirman's one step transition probabilities might be expressed
in the following form \cite{Kirman1993QJE},
\begin{eqnarray}
p( X \rightarrow X+1) &=& (N-X) \left(\sigma_1 + h X \right) \Delta t ,\label{eq:pp}\\
p( X \rightarrow X-1) &=& X \left(\sigma_2 + h [N-X] \right) \Delta t ,\label{eq:pm}
\end{eqnarray}
where $X$ is a number of agents exploiting the chosen trading
strategy (the one used to describe system state), while $N$ is
a total number of agents in the system (thus the other trading
strategy is used by the $N-X$ agents). In the above the original
Kirman's approach was extended by introducing fixed event time
scale $\Delta t$ by replacing the original models individual decision
$\varepsilon_i \rightarrow \sigma_i \Delta t$ and herding
$(1-\delta) \rightarrow h \Delta t$ parameters. Later we will need a
more general assumption that parameters $\sigma$ and $h$ may
depend on $X$ and $N$, but for now we omit it.

Note that the transition probabilities (\ref{eq:pp}) and (\ref{eq:pm}) describe
a scenario where the interaction among agent groups depends on the overall
number of agents in alternative state. Such a choice makes the transition rates
non-extensive, the connectivity between agent groups increases with the number
of agents $N$. The herding interactions have a global character. Opposite
scenario - extensive one will be also used further in this paper.

The lack of memory of the agents is the crucial assumption to formalize
the population dynamics as a Markov process. Furthermore to describe
the aforementioned dynamics in a continuous time we will need to
obtain the transition rates, transition probabilities per unit time, which for
continuous $x=X/N$ may be expressed as
\begin{eqnarray}
\pi^{+}(x) &=& (1-x) \left(\frac{\sigma_1}{N} + h x \right) ,\label{eq:pip}\\
\pi^{-}(x) &=& x \left(\frac{\sigma_2}{N} + h [1-x] \right) .\label{eq:pim}
\end{eqnarray}
Here the large number of agents $N$ is assumed to ensure the continuity
of variable $x$, which expresses the fraction of agents using the selected
trading strategy, $X$. Relation between the discrete transition probabilities,
(\ref{eq:pp}) and (\ref{eq:pm}), and continuous transition rates, (\ref{eq:pip})
and (\ref{eq:pim}), should be evident:
\begin{equation}
p ( X \rightarrow X \pm 1) = N^2 \pi^{\pm}(x) \Delta t .
\end{equation}

One can compactly express the Master equation for the system state
probability density function, $\omega(x,t)$, by using one step operators
${\bf E}$ and ${\bf E^{-1}}$ (see \cite{VanKampen1992NorthHolland}
for a details on this formalism) as
\begin{eqnarray}
\partial_t \omega(x, t) =& N^2 \left\{ ({\bf E}-1)[\pi^{-}(x) \omega(x,t)] + ({\bf E^{-1}}-1)[\pi^{+}(x) \omega(x,t)] \right\} . \label{eq:masterExpansion}
\end{eqnarray}
By expanding ${\bf E}$ and ${\bf E^{-1}}$ using the Taylor expansion
(up to the second term) we arrive at the approximation of the Master
equation
\begin{eqnarray}
\partial_t \omega(x, t) =& -N \partial_x [\{\pi^{+}(x)-\pi^{-}(x)\} \omega(x,t)] + \frac{1}{2} \partial^2_x [\{\pi^{+}(x) + \pi^{-}(x)\} \omega(x,t)]  .
\end{eqnarray}
By introducing custom functions
\begin{eqnarray}
A(x) &=& N \{\pi^{+}(x)-\pi^{-}(x)\} = \sigma_1 (1-x) - \sigma_2 x ,\\
D(x) &=& \pi^{+}(x)+\pi^{-}(x) = 2 h x (1-x) + \frac{\sigma_1}{N} (1-x) + \frac{\sigma_2}{N} x ,
\end{eqnarray}
one can make sure that the (\ref{eq:masterExpansion}) is actually
a Fokker-Planck equation:
\begin{equation}
\partial_t \omega(x, t) = - \partial_x [A(x) \omega(x,t)] + \frac{1}{2} \partial^2_x [D(x) \omega(x,t)]  .
\end{equation}
Note that in the limit of large $N$ one can neglect individual behavior
terms in the $D(x)$. The above Fokker-Planck equation was first
derived in a slightly different manner in the \cite{Alfarano2005CompEco}.

It is known (for details see \cite{Gardiner1997Springer}) that the Fokker-Planck
equation can be rewritten as Langevin equation, or in other words stochastic
differential equation,
\begin{eqnarray}
\mathrm{d} x &=& A(x) \mathrm{d} t + \sqrt{D(x)} \mathrm{d} W = [ \sigma_1 (1-x) - \sigma_2 x ] \mathrm{d} t + \sqrt{2 h x (1-x)} \mathrm{d} W , \label{eq:popSDE}
\end{eqnarray}
here $W$ stands for Wiener process. This step was also present in
the \cite{Alfarano2005CompEco}.

\begin{figure*}[!t]
\centerline{\subfloat{\includegraphics[width=2.5in]{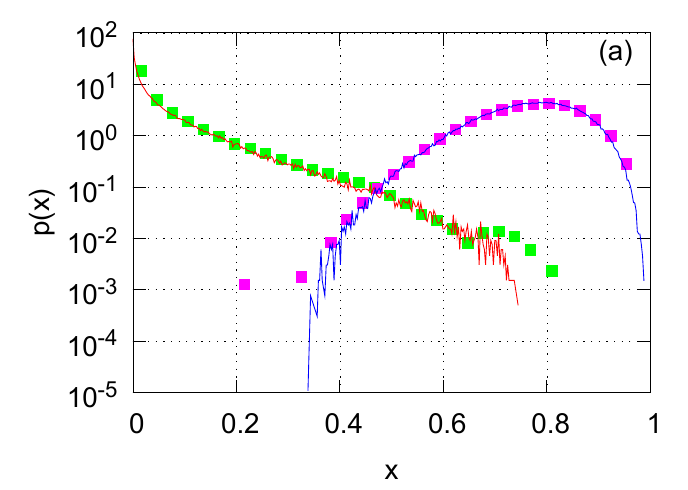}}
\hfil
\subfloat{\includegraphics[width=2.5in]{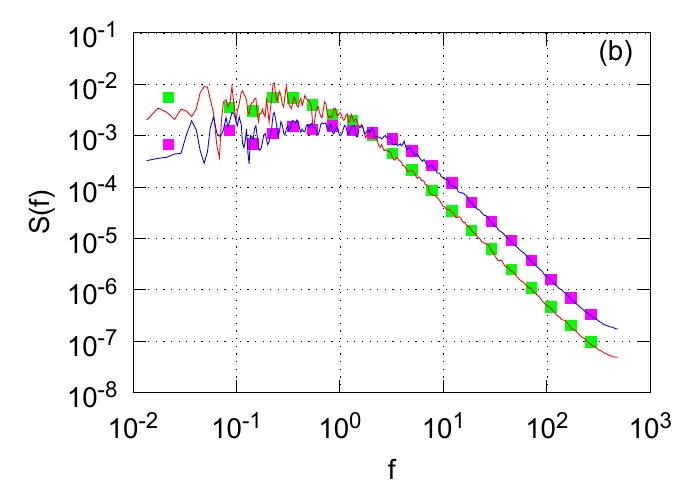}}}
\caption{Agreement between statistical properties of population fraction, $x$, (a)
probability density function and (b) power spectral density,
obtained from stochastic (red and blue curves) and agent-based
(green and magenta squares) models. Two qualitatively different model
phases are shown: red curve and green squares correspond to herding
dominant model phase ($\sigma_1=\sigma_2=0.2$, $h=5$), while blue curve
and magenta squares correspond to individual behavior dominant
model phase ($\sigma_1=\sigma_2=16$, $h=5$). Agent based model results
obtained with $N=100$.}
\label{fig:sdeAbmPopKirman}
\end{figure*}

In Fig. \ref{fig:sdeAbmPopKirman} we show that the statistical
properties obtained from the agent-based model, defined by
transition probabilities (\ref{eq:pp}) and (\ref{eq:pm}), match statistical properties
of the solutions of (\ref{eq:popSDE}). Thus the approximations done
while deriving the Langevin equation for population fraction are valid.
Interestingly enough we have obtained agreement with not so high number
of agents - $N=100$.

Note that the method used to derive Eq. (\ref{eq:popSDE}) gives us
an opportunity to consider parameters $\sigma_1$, $\sigma_2$, $h$
dependent on the variable $x$ and $N$. We will need this generalization in
the further elaboration on various applications. From our point of view,
the general form of SDE (\ref{eq:popSDE}) derived from the very basic
agent-based herding model provides a wide choice of opportunities
in consistent micro and macro modeling of complex social systems.

\section{Agent based model for the Bass Diffusion}
\label{sec:bass}

The Bass Diffusion model is a tool to forecast the diffusion rate of
new products or technologies \cite{Bass1969ManSci}. Mathematically
it is formulated as an ordinary differential equation
\begin{eqnarray}
\partial_t X(t) = [ N - X(t) ] \left[\sigma + \frac{h}{N} X(t) \right] , X(0) = 0 .
\label{eq:bass}
\end{eqnarray}
where $X(t)$ denotes the number of consumers at time $t$, $N$ can
be seen as the market potential, being a starting number of the potential
consumers (agents), $\sigma$ is the coefficient of innovation, the likelihood
of an individual to adopt the product due to influence by the commercials
or similar external sources, $h$ is the coefficient of imitation, a measure
of likelihood that an individual will adopt the product due to influence by
other people who already adopted the product. This nonlinear differential
equation serves as a macroscopic description of new product adoption
by customers widely used in business planning \cite{Mahajan1993NorthHolland}.

The agent-based approach to the same problem is related with modeling of
product adoption by individual users, or agents. One can simulate diffusion
process using computers, where individual decisions of adoption occur with
specific adoption probability affected by the other individuals in the
neighborhood. It is easy to show that Bass diffusion process is a unidirectional
case of the Kirman's herding model \cite{Kirman1993QJE}. Indeed,
let us define $x(t)$ in the same way as in previous section $x(t)=X(t)/N$,
then the potential users will adopt the product at the same rate as
in Kirman's model agents switch from one state to another
\begin{eqnarray}
\pi^{+}(x) &=& (1-x) \left(\frac{\sigma}{N} + \frac{h}{N} x
\right). \label{eq:pibass} \\
\pi^{-}(x) & = & 0 \label{eq:pmbass}.
\end{eqnarray}
The form of (\ref{eq:pmbass}) should be self explanatory - in case of the
product diffusion agent should not be allow to withdraw from the consumer
state, thus this transition probability should be forced to equal zero.

The mathematical form of (\ref{eq:pibass}) is not as evident, note that we
have substituted $h$ with $\frac{h}{N}$ (compare with the original model
transition probability (\ref{eq:pip})), and needs further discussion.
Mathematically this substitution can be backed by the need for the stochastic
term to become negligible in the limit of large $N$. In the modeled market terms
this substitution means an introduction of the interaction locality - namely it is
an assumption that each individual communicates only with his local partners
(epidemic case).

One can compare the expression of the transition probability, (\ref{eq:pibass}),
with the adoption probabilities of the Linear and GLM models of Bass Diffusion
discussed in \cite{Fibich2010NYU}. The match in expressions is clear in the small
time step limit, $\Delta t \rightarrow 0$.

In case of the transition rates (\ref{eq:pibass}) and (\ref{eq:pmbass})
the macroscopic description functions, namely drift, $A(x)$, and diffusion,
$D(x)$, become
\begin{eqnarray}
A(x) &=& N \pi^{+}(x)= (1-x) \left(\sigma + h x
\right) ,\\
D(x) &=& \pi^{+}(x)= \frac{(1-x)}{N} \left(\sigma + h x \right).
\end{eqnarray}
In the large market potential limit, $N \gg 1$, $D(x)$ becomes negligible and
thus one can consider the obtained equation to be equivalent to
the Bass Diffusion ordinary differential equation (\ref{eq:bass})
instead of the stochastic differential equation. This serves as a proof that Bass
Diffusion is an unidirectional epidemic case of Kirman's herding model. Though this
simple relation looks straightforward, we derive it and confirm by numerical simulations
in fairly original way.

\begin{figure*}[!t]
\centerline{ \subfloat{\includegraphics[width=2.5in]{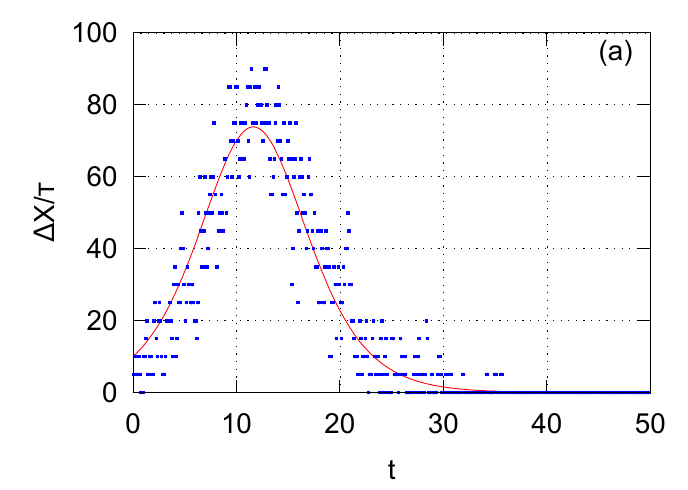}}
\hfil \subfloat{\includegraphics[width=2.5in]{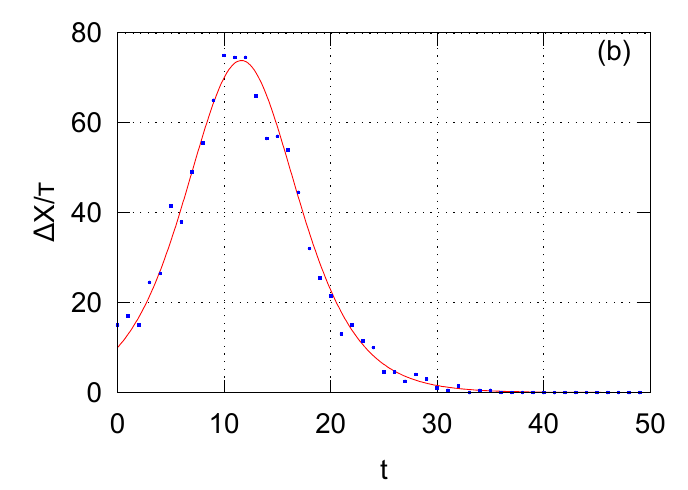}} }
\centerline{ \subfloat{\includegraphics[width=2.5in]{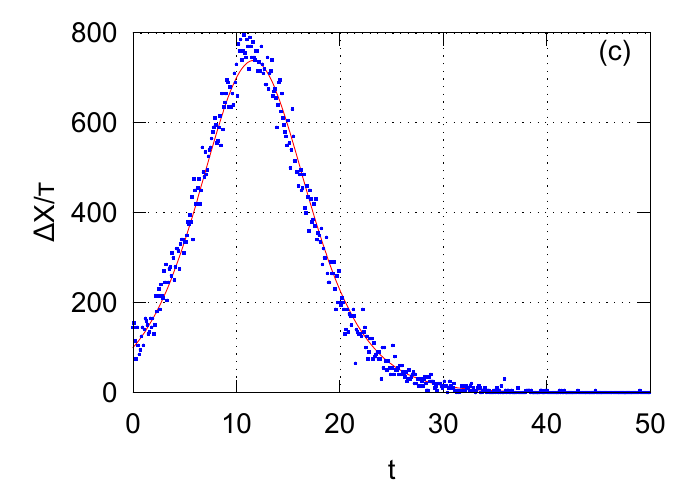}}
\hfil \subfloat{\includegraphics[width=2.5in]{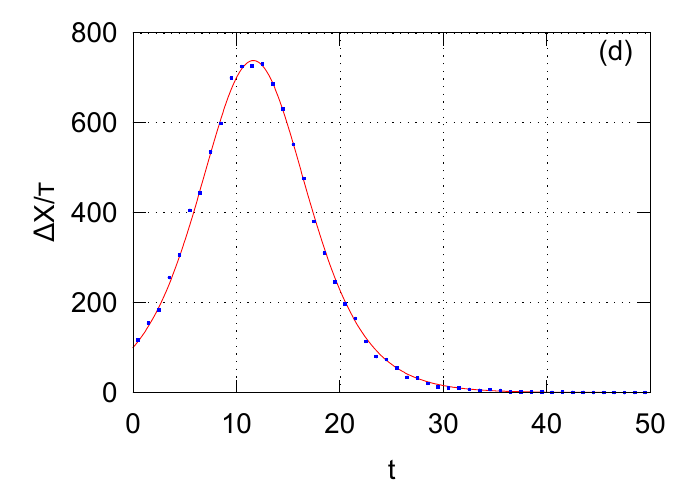}} }
\caption{Comparison of the product adoption per observation interval, $\Delta X/\tau$ versus
$t$, obtained from the macroscopic description by the Bass Diffusion model,
(\ref{eq:bass}), (red line) and the microscopic description using the
unidirectional Kirman's model, (\ref{eq:pibass}), (blue points). The models
tend to converge when time window, $\tau$, or market potential, $N$,
become larger: (a) $N=1000$, $\tau=0.1$; (b) $N=1000$, $\tau=1$;
(c) $N=10000$, $\tau=0.1$; (d) $N=10000$, $\tau=1$. Other model
parameters were the same for all subfigures and were as follows
$\sigma = 0.01$, $h = 0.275$.} \label{fig:bassAbmSde}
\end{figure*}

In Figure \ref{fig:bassAbmSde} we demonstrate the correspondence
between the Bass Diffusion model (macroscopic description) and
unidirectional Kirman's herding model (microscopic description).
Both, agent-based and continuous, descriptions of the product adoption,
$\Delta X$, converge while the market potential, $N$, or the selected
observation time interval, $\tau$, become larger.

\section{Nonlinear stochastic differential equation as a model of the financial markets}
\label{sec:financialmark}
Earlier we have introduced a class of non-linear SDEs providing time series with power-law statistics,
and most notably reproducing $1/f^\beta$ spectral density,
\cite{Gontis2004PhysA343,Kaulakys2005PhysRevE,Ruseckas2011EPL}. The
general form of the proposed class of Ito SDEs is
\begin{equation}
\mathrm{d}
y=\left(\eta-\frac{\lambda}{2}\right)y^{2\eta-1} \mathrm{d} t_s
+ y^{\eta}\mathrm{d} W_s , \label{eq:sdef}
\end{equation}
here $y$ is the stochastic process exhibiting power-law
statistics, $\eta$ is the power-law exponent of the
multiplicative noise, while $\lambda$ defines the exponent of
power-law probability density function (PDF), and $W$ is a
Wiener process (the Brownian motion). Note that SDE (\ref{eq:sdef})
is defined in the scaled time, $t_s = \sigma_t^2 t$, where $\sigma_t^2$
is the scaling parameter. Empirically we have determined that
$\sigma_t^2=1/6 \cdot 10^{-5} \mathrm{s}^{-1}$ is appropriate in
terms of the return model proposed in \cite{Gontis2010PhysA}.

From the SDE (\ref{eq:sdef})
follows that the stationary probability density function (PDF) of
this stochastic process is power-law, $p_{0}(y)\sim y^{-\lambda}$,
with the exponent $\lambda$ \cite{Gardiner1997Springer}. While in
Refs. \cite{Kaulakys2006PhysA} and later more precisely in
\cite{Ruseckas2010PhysRevE} it was shown that the time series
obtained while solving SDE (\ref{eq:sdef}) have power-law spectral density
\begin{equation}
S(f)\sim \frac{1}{f^{\beta}} ,
\qquad\beta=1+\frac{\lambda-3}{2(\eta-1)} .
\end{equation}
Note that exponent of spectral density, $\beta$, is defined only for $\eta \neq 1$. In case of
$\eta =1$ the SDE (\ref{eq:sdef}) becomes identical to the geometric
Brownian motion.

Power law statistics of the signal $y$ obtained by solving SDE (\ref{eq:sdef})
and exponents $\lambda$, $\beta$ are defined for large $y$ values. Thus one
has to introduce the diffusion restriction terms in the limit of small $y$ values
when attempting to solve SDE (\ref{eq:sdef}) or applying it in a
stochastic modeling. There is a wide choice of restriction mechanisms adjustable
to the needs of real systems with negligible influence on the power law exponents.
We have introduced a term of additive noise while attempting to model the absolute
return \cite{Gontis2010PhysA}
\begin{eqnarray}
\mathrm{d}
y=\left(\eta-\frac{\lambda}{2}\right)(1+y^2)^{\eta-1}y\mathrm{d}
t_s +(1+y^2)^{\frac{\eta}{2}}\mathrm{d} W_s. \label{eq:power2}
\end{eqnarray}
 In such case the stationary probability density function of the SDE (\ref{eq:sdef})
is a $q$-Gaussian (see \cite{Gontis2010Sciyo,Gontis2010PhysA})
\begin{equation}
P_{\lambda}(y)=\frac{\Gamma(\lambda/2)}{\sqrt{\pi}\Gamma(\lambda/2-1/2)}
\left(\frac{1}{1+y^2}\right)^{\frac{\lambda}{2}} .
\label{eq:qGaussian2}
\end{equation}

While modeling the trading activity \cite{Gontis2008PhysA} we have used
the exponential diffusion restriction for small values of variable
$y\simeq y_{\mathrm{min}}$
\begin{eqnarray}
\mathrm{d} y&=&\left[\eta-\frac{1}{2}\lambda+\frac{m}{2}
\left(\frac{y_{\mathrm{min}}^{m}}{y^{m}}
\right)\right]y^{2\eta-1}\mathrm{d} t+  y^{\eta}\mathrm{d} W.
\label{eq:sde-restricted}
\end{eqnarray}
Equation (\ref{eq:sde-restricted}) has a very general form,
which includes the well known models applicable to financial markets such
as the \emph{Cox-Ingersoll-Ross} (CIR) process or the
\emph{Constant Elasticity of Variance} (CEV) process \cite{Jeanblanc2009Springer}
\begin{equation}
\mathrm{d} y=\mu y\mathrm{d} t+y^{\eta}\mathrm{d} W ,
\end{equation}
where $\mu=(\eta-1)y_{\mathrm{min}}^{2(\eta-1)}$,
as a less general cases of the SDE (\ref{eq:sde-restricted}).

The class of equations based on SDE (\ref{eq:sdef}) gives only a general idea how
to model power-law statistics of trading activity and return in the financial markets.
The problem is to determine the parameter set $\lambda$ and $\eta$ in
a way giving the empirical values for the $\lambda$ and $\beta$.
The task becomes even more complicated if one considers the more sophisticated
behavior of the spectral density - power spectral densities have not
one, but two power-law regions with different values of $\beta$. In the series of
papers \cite{Gontis2010PhysA,Kaulakys2006PhysA,Gontis2008PhysA,Gontis2006JStatMech} we have
shown that trading activity and return can be modeled by a more sophisticated
version of the SDE than (\ref{eq:sdef}) now including the two powers of the noise
multiplicativity. In the case of return instead of Eq. (\ref{eq:power2}) one should
use
\begin{eqnarray}
\mathrm{d} y&=&\left(\eta-\frac{\lambda}{2}\right)
\frac{(1+y^2)^{\eta-1}}{(\epsilon \sqrt{1+y^2}+1)^2}y\mathrm{d}
t_s +\frac{(1+y^2)^{\frac{\eta}{2}}}{\epsilon \sqrt{1+y^2}+1}\mathrm{d}
W_s, \label{eq:return2}
\end{eqnarray}
here $\epsilon$ divides the area of diffusion into the two different
noise multiplicativity regions to ensure the spectral density of $|y|$ with two
power law exponents.

The proposed form of the SDE enables reproduction of the main statistical
properties of the return observed in the financial markets. Similarly one can
deal with a more sophisticated model for the trading activity \cite{Gontis2008PhysA}.
This provides an approach to the financial markets with behavior dependent on
the level of activity and exhibiting two stages: calm and excited. Equation
(\ref{eq:return2}) models the stochastic return $y$ with two power-law statistics,
namely the probability density function and the power spectral density,
reproducing the empirical power law exponents of the return in the financial markets.

\section{Kirman's model as a microscopic approach to the financial markets}
\label{sec:kirmanfinance}
The drawback of the stochastic models is a lack of direct insights into the microscopic
nature of replicated dynamics. Bridging between microscopic and macroscopic
approaches is needed for better grounding of stochastic modeling.

Top-down approach, namely starting from the stochastic modeling and moving
towards the agent-based models, seems to be a very formidable task, as the
macro-behavior of complex system can not be understood as a simple superposition
of varying micro-behaviors. While in the case of sophisticated agent-based models
\cite{Cristelli2010Fermi} bottom-up approach provides too many opportunities.
But there is selection of rather simple agent-based models (ex. \cite{Kirman1993QJE}),
whose stochastic treatment can be directly obtained from the microscopic description
\cite{Alfarano2005CompEco}.

Here we consider an opportunity to generalize Kirman's ant colony model \cite{Kirman1993QJE}
with the intention to modify its microscopic approach to the financial market modeling
\cite{Alfarano2005CompEco} reproducing the main stylized facts of this complex system.
In the Section \ref{sec:kirman} we have already introduced Kirman's ant colony
model, proposed its generalization and derived stochastic model for the two state
population dynamics.

As Kirman's model considers the two available agent states one must define two types
of agents acting inside the market in order to relate Kirman's model to financial markets.
Currently, the most common choice is assuming that agents can be either fundamentalists
or noise traders \cite{Cristelli2010Fermi}.

Fundamentalists are assumed to have the fundamental knowledge about the market,
which is assumed to be quantified by the so-called fundamental price, $P_f(t)$, of the
traded stock. By having this knowledge they can make long term forecasts on a notion
that infinitely long under-evaluation or over-evaluation of the stock is impossible - the
market in some point in the future will have to set a fair price on the stock. Thus their
excess demand, which is shaped by their long term expectations, is given by
\cite{Alfarano2005CompEco}
\begin{equation}
D_f(t) = N_f(t) \ln \frac{P_f(t)}{P(t)} ,
\end{equation}
here $N_f(t)$ is a number of fundamentalists inside the market and $P(t)$ is a current
market price. As long term investors fundamentalists assume that $P(t)$ will converge
towards $P_f(t)$ at least in a long run. Therefore if $P_f(t) > P(t)$, fundamentalists will
expect that $P(t)$ will grow in future and consequently they will buy the stock ($D_f(t)>0$).
In the opposite case, $P_f(t) < P(t)$, they will expect decrease of $P(t)$ and for this reason
they will sell the stock ($D_f(t)<0$).

The other group, the noise traders are investors who attempt estimate the stocks future
price based on its recent movements. As there is a wide selection of technical trading
strategies, which are used to analyze stocks price movements, one can simply assume that
the average noise traders demand is based on their mood, $\xi(t)$, \cite{Alfarano2005CompEco}
\begin{equation}
D_c(t) = r_0 N_c(t) \xi(t) ,
\end{equation}
here $D_c(t)$ is a total excess demand of noise trader group, $N_c(t)$ is a number of
noise traders inside the market and $r_0$ can be seen as a relative noise trader
impact factor.

Price and, later after a brief derivation, return can be introduced into the model by
applying the Walrassian scenario. One can assume that trading in the market is cleared
instantaneously to set
a price, which would stabilize the market demand for a given moment. Thus the sum of all
groups' excess demands should equal zero:
\begin{eqnarray}
D_f(t) + D_c(t) &=&  N_f(t) \ln \frac{P_f(t)}{P(t)}  + r_0 N_c(t) \xi(t) = 0 ,\\
P(t) &=& P_f(t) \exp \left[ r_0 \frac{N_c(t)}{N_f(t)} \xi(t) \right] ,
\end{eqnarray}
where without loosing generality one can assume that fundamental price remains constant, $P_f(t) = P_f$.

Consequently the return, which is defined as logarithmic change of price, in the selected
time window $T$ is given by:
\begin{eqnarray}
r(t)&=&\ln P(t) - \ln P(t-T) = r_0 \left[ \frac{x(t)}{1-x(t)} \xi(t) -  \frac{x(t-T)}{1-x(t-T)} \xi(t-T) \right] , \label{eq:defrnonadiab}
\end{eqnarray}
where we have set that $\frac{N_c(t)}{N}=x$ and $\frac{N_f(t)}{N}=1-x$ according
to the notation introduced in Section \ref{sec:kirman}. Alfarano et al.
\cite{Alfarano2005CompEco} simplified the above by assuming that $x(t)$ is significantly
slower process than $\xi(t)$, obtaining adiabatic approximation of the return
\begin{equation}
r(t) = r_0 \frac{x(t)}{1-x(t)} \zeta(t) ,
\end{equation}
where $\zeta(t) = \xi(t)-\xi(t-T)$. If $\zeta(t)$ is modeled using spin-noise model,
as in \cite{Alfarano2005CompEco}, then the middle term, $\frac{x(t)}{1-x(t)}$,
can be seen as an absolute return.

\begin{figure*}[!t]
\centerline{\subfloat{\includegraphics[width=2.5in]{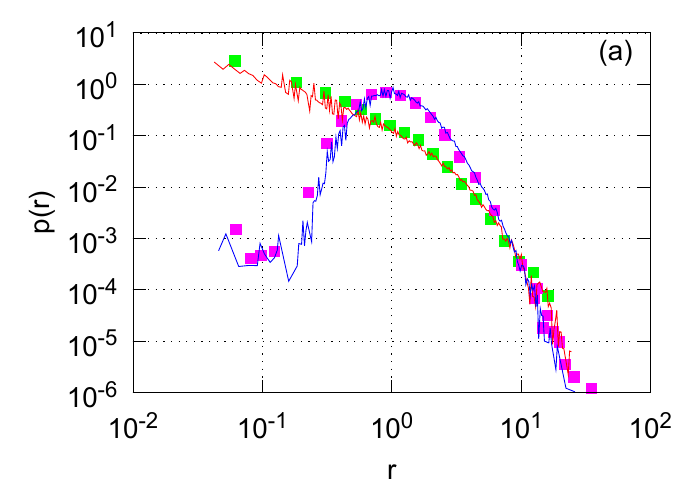}}
\hfil
\subfloat{\includegraphics[width=2.5in]{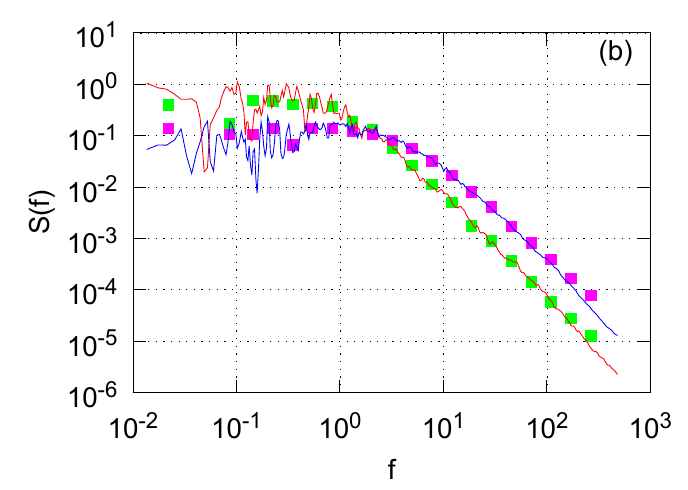}}}
\caption{Agreement between statistical properties of $y$, (a)
probability density function and (b) power spectral density,
obtained from the stochastic (red and blue curves) and agent-based
(green and magenta squares) models. Two qualitatively different model
phases are shown: red curve and green squares correspond to herding
dominant model phase ($\sigma_1=\sigma_2=0.2$, $h=5$), while blue curve
and magenta squares correspond to individual behavior dominant
model phase ($\sigma_1=\sigma_2=16$, $h=5$). Agent based model results
obtained with $N=100$.}
\label{fig:sdeAbmKirman}
\end{figure*}

Using Ito formula for variable substitution \cite{Gardiner1997Springer} in SDE (\ref{eq:popSDE}) we obtain
nonlinear SDE for the $y(t)=\frac{x(t)}{1-x(t)}$
\begin{eqnarray}
\mathrm{d} y &=& ( \sigma_1 - y [ \sigma_2 - 2 h] ) (1+y) \mathrm{d} t +  \sqrt{2 h y}(1+y) \mathrm{d} W . \label{eq:ysde}
\end{eqnarray}
Agreement between the agent-based Kirman's model applied towards financial markets
using the ideas discussed above and the new stochastic model for $y$, (\ref{eq:ysde}),
is demonstrated in Fig. \ref{fig:sdeAbmKirman}.

Note once again that the actual derivation, and thus, the final outcome,
does not change even if $\sigma_1$, $\sigma_2$ or $h$ are the
functions of either $x$ or $y$. Therefore, one can further study the
possibilities of the obtained stochastic model, (\ref{eq:ysde}), by
checking different scenarios of $\sigma_1$, $\sigma_2$ or $h$ being
functions of either $x$ or $y$. Nevertheless, the most natural way
is to introduce a custom function $\tau(y)$ to adjust the inter-event
time according to the system state. From the financial market point of
view this can be seen as introduction of variability of trading activity
based on the return.

We have chosen the case when $h$ and $\sigma_2$ are functions
of $y$, namely we make the substitutions,
$\sigma_2 \rightarrow \frac{\sigma_2}{\tau(y)}$ and
$h \rightarrow \frac{h}{\tau(y)}$, in the Kirman's model transition
probabilities, (\ref{eq:pp}) and (\ref{eq:pm}), and stochastic
model for $y$, (\ref{eq:ysde}). To further simplify the model we
can introduce scaled time, $t_s = h t$, and make related model
parameter transformations, $\varepsilon_i = \frac{\sigma_i}{h}$.
By making these substitutions we arrive at
\begin{eqnarray}
\mathrm{d} y &=& \left[ \varepsilon_1 + y \frac{2 - \varepsilon_2}{\tau(y)}
\right] (1+y) \mathrm{d} t_s +\sqrt{\frac{2 y}{\tau(y)}} (1+y)
\mathrm{d} W_s , \label{eq:ytaucommon}
\end{eqnarray}
where $W_s$ is appropriately scaled Wiener process. Note that we
left $\sigma_1$, and consequently $\varepsilon_1$, independent of
$y$ on purpose as one could argue that individual behavior of
fundamentalist trader should not depend on the observed returns
as he is a long term investor uninterested in the momentary fluctuations
of the market mood.

Note that absolute return, $y$, defined in Eqs. (\ref{eq:ysde}) and (\ref{eq:ytaucommon}), serve as a
measure of volatility in the financial markets. It is known that volatility
has long-range memory and correlates with trading activity and has
probability density function with power law tail \cite{Cont2001RQUF}.
We are particularly interested in the case of $\tau(y)=y^{-\alpha}$.
This selection is defined by the fact that trading activity has positive
correlation with volatility and the class of SDE (\ref{eq:sdef}) is
invariant regarding power-law variable transformation, see
\cite{Ruseckas2011EPL}. In such case the obtained stochastic
differential equation, Eq. (\ref{eq:ytaucommon}), in the limit of $y \gg 1$ is
very similar to the stochastic models discussed in the Section
\ref{sec:financialmark}.

In the aforementioned limit of $y$, $y \gg 1$, we can consider only
the highest powers in Eq. (\ref{eq:ytaucommon}). In such case Eq.
(\ref{eq:ytaucommon}) is reduce to the
\begin{equation}
\mathrm{d} y = (2-\varepsilon_2) y^{2+\alpha} \mathrm{d} t_s + \sqrt{2 y^{3+\alpha}} \mathrm{d} W_s . \label{eq:sdeylarge}
\end{equation}
The direct comparison of Eqs. (\ref{eq:sdef}) and
(\ref{eq:sdeylarge}) yields:
\begin{equation}
\eta = \frac{3+\alpha}{2}, \quad \lambda = \varepsilon_2 + \alpha +1 . \label{eq:etadef}
\end{equation}
Consequently we expect that the stochastic process $y$ defined
by Eq. (\ref{eq:sdeylarge}) will have the power law stationary probability
density function,
\begin{equation}
p(y) \sim y^{-\varepsilon_2 -\alpha-1} , \label{eq:pdfJulius}
\end{equation}
and also a power law spectral density,
\begin{equation}
S(f) \sim \frac{1}{f^\beta}, \quad \beta = 1 + \frac{\varepsilon_2 + \alpha - 2}{1+\alpha} , \label{eq:specJulius}
\end{equation}
where we have used the relation between model parameters, Eq. (\ref{eq:etadef}).

While if we linearize drift function of Eq. (\ref{eq:ysde}) with
the respect to the absolute return, $y$, namely set $\varepsilon_2 = 2$,
we would obtain a stochastic differential equation (once again in the limit $y \gg 1$)
\begin{equation}
\mathrm{d} y = \varepsilon_1 y \mathrm{d} t_s + \sqrt{2 y^{3+\alpha}} \mathrm{d} W_s . \label{eq:sdeylarge2}
\end{equation}
similar to the generalized CEV process \cite{Jeanblanc2009Springer,Reimann2011PhysA},
which was considered in \cite{Reimann2011PhysA},
\begin{equation}
\mathrm{d} y = a y \mathrm{d} t + b y^\eta \mathrm{d} W. \label{eq:cev}
\end{equation}
In \cite{Ruseckas2011EPL} the latter was noted to be a special case
of Eq. (\ref{eq:sde-restricted}) with exponential restriction of diffusion
applied. The comparison with this special case is important on its
own as this equation generalizes some stochastic models used in
risk management. Theoretical prediction of PDF
and spectral density for $y$ defined by Eq. (\ref{eq:sdeylarge2}), is given
by \cite{Reimann2011PhysA}
\begin{eqnarray}
& p(y) \sim y^{-3-\alpha} , \label{eq:pdfcev}\\
& S(f) \sim \frac{1}{f^\beta}, \quad \beta=1+\frac{\alpha}{1+\alpha} , \label{eq:speccev}
\end{eqnarray}
where we have used the previously obtained relation between model
parameters, Eq. (\ref{eq:etadef}).

\begin{figure*}[!t]
\centerline{
\subfloat{\includegraphics[width=2.5in]{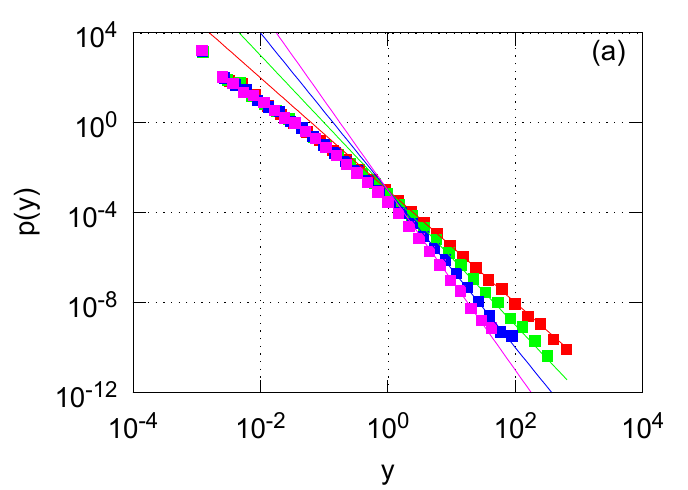}}
\hfil \subfloat{\includegraphics[width=2.5in]{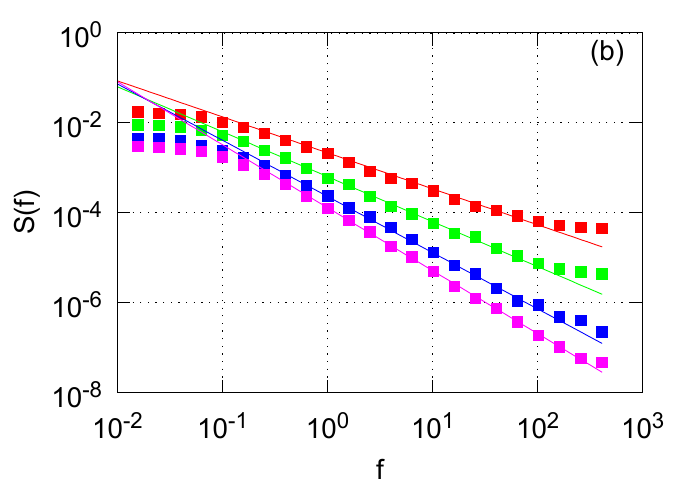}}
} \centerline{
\subfloat{\includegraphics[width=2.5in]{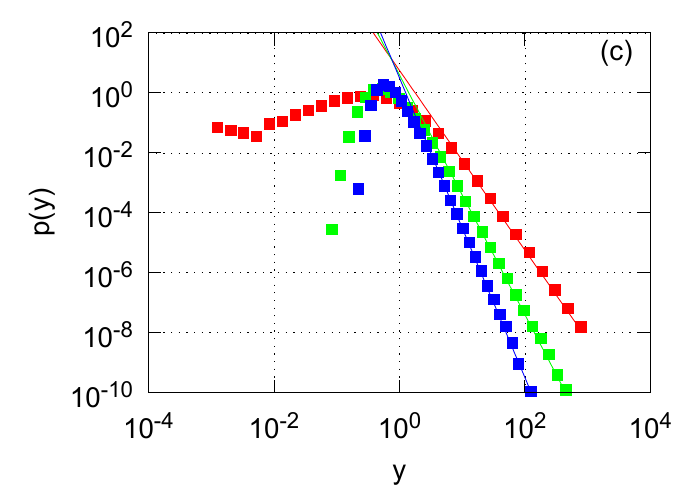}}
\hfil \subfloat{\includegraphics[width=2.5in]{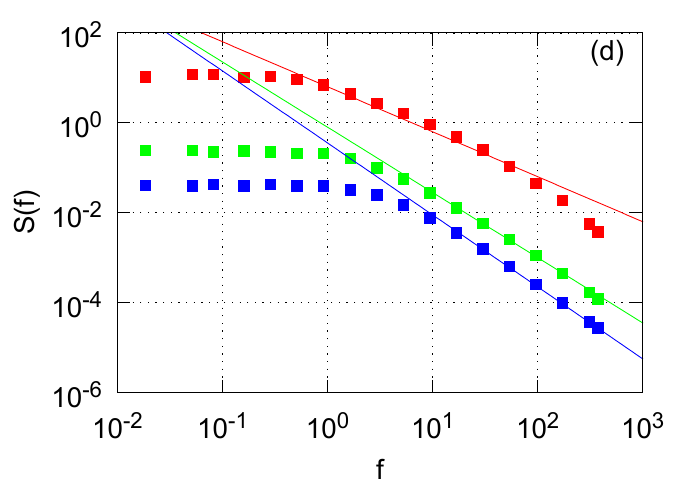}}
}
\caption{Statistical properties, (a) and (c) - probability density function,
(b) and (d) - spectral density, of the time series obtained by solving Eq. (\ref{eq:ytaucommon}) (colored squares).
Fits are provided by the theoretical predictions made in this section, (a) and (b) are fitted by using
(\ref{eq:pdfJulius}) and (\ref{eq:specJulius}), (c) and (d) are fitted by using (\ref{eq:pdfcev}) and
(\ref{eq:speccev}), (curves of corresponding colors). Model parameters for the (a) and (b) were set as follows: $\alpha=1$,
$\varepsilon_1 = 0$, $\varepsilon_2=0.5$ (red squares), $1$ (green squares),
$1.5$ (blue squares) and $2$ (magenta squares). Model parameters for the (c) and (d)
were set as follows: $\varepsilon_1 = \varepsilon_2 = 2$, $\alpha=0$ (red squares), $1$
(green squares) and $2$ (blue squares).}
\label{fig:compCEV}
\end{figure*}

In the Figure \ref{fig:compCEV} we show that the theoretical predictions
discussed in this section are valid and that they enable the reproduction
of different spectral densities and probability density functions.

Note that while the stochastic model based on herding behavior of agents
appears to be too crude to reproduce statistical properties of financial markets
in such details as the stochastic model driven by the Eq. (\ref{eq:return2}), which is
heavily based on the empirical research, it  contains very important long range
power law statistics of the absolute return. Obtained equations are very similar
to some general stochastic models of the financial markets \cite{Ruseckas2010PhysRevE,
Reimann2011PhysA} and thus, in future development might be able to serve
as a microscopic justification for them and maybe for the more
sophisticated model driven by the Eq. (\ref{eq:return2}).

\begin{figure*}[!t]
\centerline{
\subfloat{\includegraphics[width=2.5in]{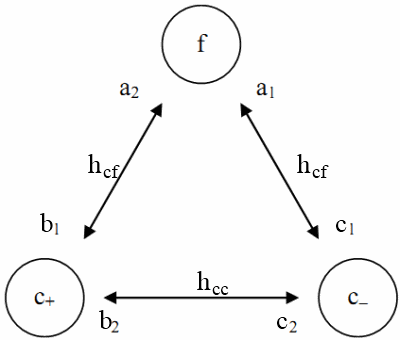}}
}
\caption{The general case of the three groups of interacting agents: $f$ - fundamentalists,
$c_{+}$ - chartists optimists, $c_{-}$ - chartists pessimists. $h_{ij}$ are herding terms,
while $a_i$, $b_i$ and $c_i$ stand for individual transitions in the direction of the arrow.}
\label{fig:3agents}
\end{figure*}

It is possible to extend agent-based model by introducing additional agent groups
or splitting old ones. Let us assume that chartist agents may disagree in their
expectations and thus divide into pessimists and optimists. Therefore it is natural
to introduce three agent groups (see Fig. \ref{fig:3agents}) interacting among themselves.
Our first attempts in this direction proves that in case of the three agent groups
(as shown in Fig. \ref{fig:3agents}), when the herding parameter $h_{cc} \gg h_{cf}$,
might confirm the expectation of a more complex behavior exhibiting fractured power
spectral density of absolute return. More detailed study of such approach in comparison
with macroscopic modeling by SDE (\ref{eq:return2}) is ongoing.

\section{Multifractal behavior of return series}
\label{sec:mfdfa}

In the last few decades it was noted that many natural phenomena have very complex
intrinsic structure, which has a very specific scaling properties. This notion was generalized
as fractal framework \cite{Feder1988Plenum}. Later it was also noted that the scaling
properties of some processes exhibit even more complex scaling behavior - namely
they appeared to have features of the multiple fractals. Few examples of such
phenomena include geoelectrical processes \cite{Telesca2005NJP}, human heartbeat
\cite{Ivanov1999Nature} and gait \cite{West2003PhysRevE}. The financial market
time series apparently are also of the multifractal nature
\cite{Ausloos2002CompPhysComm,Peters1994Willey,Kwapien2005PhysA}.

There are few established methods to detect multifractal time series and two very
prominent methods. One of them is generalized height-height correlation function method
(GHHCF) and multifractal detrended analysis method (MF-DFA). In our previous approaches
\cite{Kaulakys2006WorldSc,Kononovicius2012PhysA} we have used the GHHCF method, so
let us in this contribution to rely on the MF-DFA method.

To start with the multifractal analysis of the time series, $y_k$, we have to obtain the profile of
the time series, $Y_k$:
\begin{equation}
Y_k = \sum_{j=1}^k (y_k - \langle y \rangle ) .
\end{equation}
Next we have divide the $Y_k$ series into equally sized and non overlapping segments. Thus
if our segments are of the size $s$, then we will have $N_s = \mathrm{int}(N/s)$ segments
(here $N$ is the length the series, while $\mathrm{int}(\dots)$ is a function which takes an
integer part of the argument). For the most of the segment sizes some of the data will be lost,
in order to account for it one might want to take another set of segments, but now splitting
from the end of the series.

Further, one has to determine the trends in the obtained segments. Generally this can be done using
varying polynomial fits, but linear fits in the most cases are more than enough. After the trends,
$\bar Y$, are known one has to evaluate how well the trend fits the actual series:
\begin{eqnarray}
F^2_{\nu}(s) &=& \frac{1}{s} \sum_{i=1}^{s} \left[ Y_{(\nu-1) s +i} - {\bar Y}_{\nu}(i) \right]^2 , \label{eq:mfdfaNuSmall} \\
F^2_{\nu}(s) &=& \frac{1}{s} \sum_{i=1}^{s} \left[ y_{N-(\nu-N_s) s +i} - {\bar y}_{\nu}(i) \right]^2 . \label{eq:mfdfaNuLarge}
\end{eqnarray}
The Eq. (\ref{eq:mfdfaNuSmall}) holds for segments $\nu = 1, \ldots, N_s$, while the Eq.
(\ref{eq:mfdfaNuLarge}) should applied towards segments $\nu = N_s +1, \ldots, 2 N_s$.
Finally one has to average over all segments using
\begin{equation}
F_q(s) = \left\{ \frac{1}{2 N_s} \sum_{\nu=1}^{2 N_s} \left[ F^2_{\nu}(s) \right]^{\frac{q}{2}} \right\}^{\frac{1}{q}} ,
\end{equation}
here $q$ stands for generalized coefficient, which is the one enabling us to recover
multifractal features – it is also the only difference from the original detrended fluctuation
analysis (DFA) method \cite{Kantelhardt2002PhysA}. Note that in case of $q=2$ the $F_q(s)$
is the same as the one in the original DFA method.

All that is left is to determine is the power law trend, $h(q)$, of the $F_q(s)$. These trends, $h(q)$, are
also frequently named the generalized Hurst exponents. If the Hurst exponents are different for
different $q$, which can be any real number, then the signal can be seen as multifractal. In the
opposite case or if the variation is negligible, time series can be assumed as monofractal. For
more details on the MF-DFA method see \cite{Kantelhardt2002PhysA}.

\begin{figure*}[!t]
\centerline{
\subfloat{\includegraphics[width=2.5in]{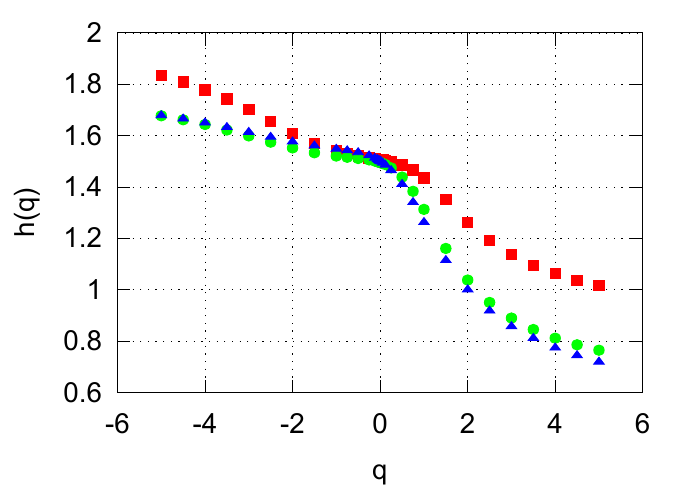}}
}
\caption{The broad spectra of Hurst exponents, $h(q)$, obtained from time series obtained
by solving (\ref{eq:ytaucommon}). The model parameters were set as follows: $\varepsilon_1=1$,
$\varepsilon_2=2-\alpha$, $\alpha=0$ (red squares), $1$ (green circles) and $2$ (blue triangles).}
\label{fig:mfHurst}
\end{figure*}

In Figure \ref{fig:mfHurst} we show that the stochastic differential equations obtained for the
modeling of financial markets and derived from the Kirman's agent-based model have broad multifractal
spectra. The curves capture a region of the Brownian motion, $h(q) \approx 1.5$, and a region of
long range memory, $h(q) \approx 1$. Note that in case of $\alpha=1$ and $\alpha=2$ (green and
blue curves) $h(2)=1$, which can be seen as a proof that the obtained time series posses
the long term correlated (have so-called long range memory)
behavior, while for $\alpha=0$ interim behavior between the Brownian motion and long range
memory is observed, $1 < h(2) < 1.5$.

\section{Statistics of bursts generated by nonlinear SDE}
\label{sec:bursts}

In the Section \ref{sec:kirmanfinance} we have shown that the herding model of return
in the financial markets leads to the class of stochastic differential equations, whose general form
is given by SDE (\ref{eq:sdef}). This class of stochastic differential equations reproduces
power law statistics, namely the probability density function and the spectral density,
of return and trading activity in the financial markets. The burst statistics of the financial markets
are also very important for the risk management and would serve as an additional
criteria to determine the model consistency.  In this section we provide
some initial results of burst statistics generated by the SDE (\ref{eq:sdef}).

\begin{figure*}[!t]
\centerline{
\subfloat{\includegraphics[width=2.5in]{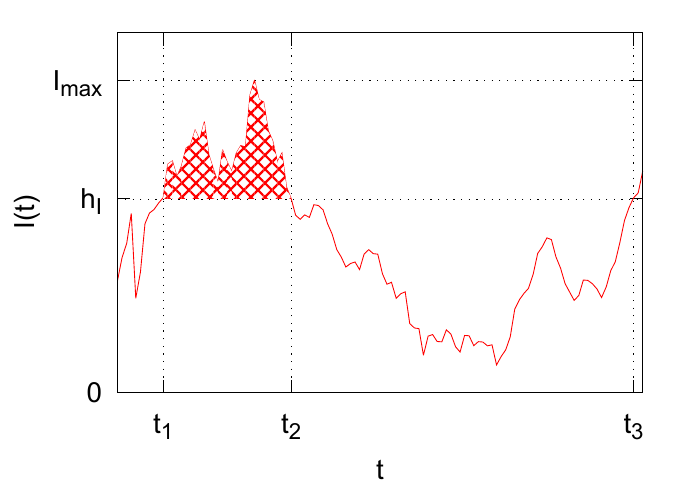}}
}
\caption{Time series exhibiting bursty behavior, $I(t)$. Here $h_I$ is
the threshold value, above which bursts are detected, $t_i$ is the
three visible threshold passage events, $I_{max}$ is the
highlighted burst's peak value. Burst duration we define as: $T= t_2 - t_1$.} \label{fig:burstExample}
\end{figure*}

We define a burst as a part of the time series lying above the certain threshold, $h_I$.
In Figure \ref{fig:burstExample} we present an example burst of the simple bursty time
series, $I(t)$. Evidently a burst as itself can be described
by its duration, $T = t_2 - t_1$, maximum value, $I_{max}$, and burst size, which
we define as an area above the selected threshold yet bellow time series curve
(highlighted by x pattern in the Fig. \ref{fig:burstExample}), $S$.

There is a well established passage, or alternatively hitting, time framework, which is
frequently used to tackle practical problems in both mathematical finance
\cite{Jeanblanc2009Springer} and physics \cite{Gardiner1997Springer,Redner2001Cambridge}.
One can also apply this framework to understand the burst durations, $T$. Interestingly
enough we can consider the first hitting time of the stochastic process starting infinitesimally near the
hitting threshold as the burst duration itself, $T$.

Brownian motion, geometric Brownian motion and Bessel process are highly applicable
models (for examples of the application in the mathematical finance, see
\cite{Jeanblanc2009Springer}) for which hitting times statistics are known.
The Bessel process,
\begin{equation}
\mathrm{d} R = \frac{N-1}{2} \frac{\mathrm{d} t_s}{R} + \mathrm{d} W_s , \label{eq:BesselProcSDE}
\end{equation}
is one of the most interesting as some prominent mathematical finance models can
be transformed to a similar form. In order to simplify further handling of the
Bessel process it is convenient to introduce $\nu= \frac{N}{2} -1$, which is known
as the index of the Bessel process. While $N$ is also frequently retained and mentioned
as it bears an actual physical meaning - the Bessel process is an Euclidean norm,
length of the vector, of $N$-dimensional Brownian motion, which starts at the origin.
Note that for $N>1$, or alternatively $\nu>-0.5$, $R$ tends to diverge towards infinity.

In our case the Bessel process is of high interest as by using the Lamberti transform
defined as
\begin{equation}
\ell:y \mapsto z(y) = \frac{1}{(\eta-1) y^{\eta-1}} , \label{eq:varTrans}
\end{equation}
we can reduce a general class of SDE (\ref{eq:sdef}) to the Bessel process,
\begin{equation}
\mathrm{d} z = \left( \nu + \frac{1}{2} \right) \frac{\mathrm{d}
t_s}{z} + \mathrm{d} W_s , \label{eq:BesselTransSDE}
\end{equation}
with index $ \nu = \frac{\lambda - 2 \eta + 1}{2 (\eta -1)}$. The corresponding
dimension of the Brownian motion is given by
$ N = 2 (\nu+1)= \frac{\lambda-1}{\eta-1} $.

Let us assume that a burst starts at time $t_0$, with $y_0 = y(t_0)$
slightly exceeding the selected threshold, $h_y$. By definition
the burst lasts until $y(t)$ crosses $h_y$ once again, but now from the above.
Equivalently, in the terms of Bessel process the burst lasts until at
a certain time, $t$, the $z$ process crosses the boundary $h_z=\ell(h_y)$
from the below, while the starting position, $z_0 =z(t_0)$, which in the terms of
Bessel process is below the threshold, $z_0 = \ell(y_0)< h_z$.

Consequently by choosing $z_0$ arbitrarily close yet below $h_z$,
we can obtain an estimate for the burst duration, $T$, in terms of
the hitting times of the Bessel process, $\tau^{(\nu)}_{z_0,h_z}$,
\begin{eqnarray}
T &=& \tau^{(\nu)}_{z_0,h_z} = \inf_{t>t_0} \Big\{ t , \: z(t) \geq h_z \Big\} ,  0 < h_z-z_0 < \epsilon ,
\end{eqnarray}
where $\epsilon$ is an arbitrary small positive constant. As given in
\cite{Borodin2002Birkhauser}, the following holds for $0 < z_0 < h_z$
\begin{eqnarray}
\rho^{(\nu)}_{z_0,h_z}(t) &=& \frac{h_z^{\nu-2}}{z_0^\nu} \sum_{k=1}^{\infty} \left[ \frac{j_{\nu,k} J_\nu\left(\frac{z_0}{h_z} j_{\nu,k}\right)}{J_{\nu+1}(j_{\nu,k})} \exp\left(- \frac{j_{\nu,k}^2}{2 h_z^2} t\right) \right] ,
\end{eqnarray}
where $\rho_{z_0,h_z}^{(\nu)}(t)$ is a probability density function of
the hitting times at level $h_z$ of Bessel process with index $\nu$ starting
from $z_0$, $J_\nu$ is a Bessel function of the first kind of the order $\nu$, while
$j_{\nu,k}$ is a $k$-th zero of $J_\nu$.

\begin{figure*}[!t]
\centerline{
\subfloat{\includegraphics[width=2.5in]{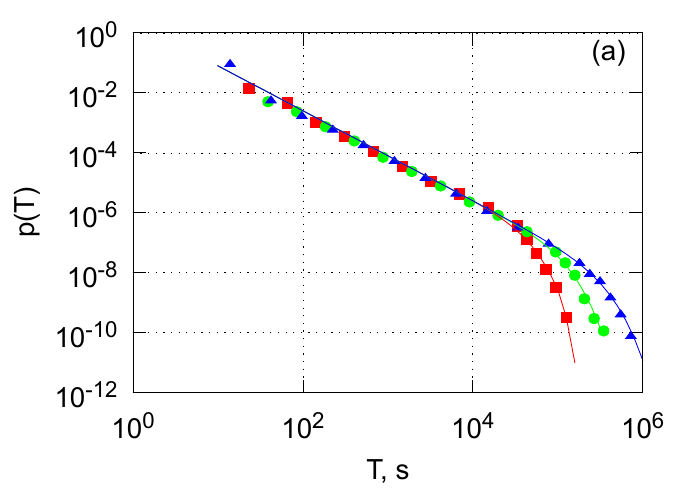}}
\hfil \subfloat{\includegraphics[width=2.5in]{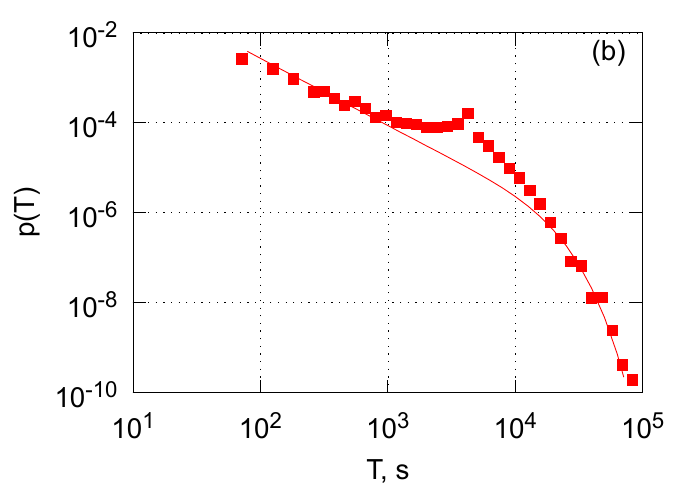}}
}
\caption{ Numerical (a) and empirical (b) PDF of burst
durations, $h_y=2$. In both subfigures numerical and empirical data is represented
by filled shapes, while fits, (\ref{eq:durationApprox}), are represented by gray curves.
Model, (\ref{eq:sdef}), parameters were set as follows: $\sigma_t^2 = 1/6 \cdot 10^{-5} \mathrm{s}^{-1}$ (in all three cases), $\lambda=4$ (in all three cases), $\eta = 2.5$ (red squares, $\nu =0$), $\eta = 2$ (green circles, $\nu =0.5$) and $\eta=1.5$ (blue triangles, $\nu =2$). Empirical data fitted by assuming that $\nu=-0.2$.}
\label{fig:burstDuration}
\end{figure*}

We have to replace $\rho^{(\nu)}_{z_0,h_z}(t)$ by density function
regarding $h_z$ to avoid the self-evident convergence of
$\rho^{(\nu)}_{z_0,h_z}(t)$ (for $t>0$) to zero, when $z_0 \rightarrow h_z$.
This is achieved introducing the probability density function $p_{h_z}^{(\nu)}(t)$
as a probability density function of the burst duration
\begin{equation}
p_{h_z}^{(\nu)}(t) = \lim_{z_0 \rightarrow h_z} \frac{\rho^{(\nu)}_{z_0,h_z}(t)}{h_z-z_0} , \label{eq:durationPrecise}
\end{equation}
where we have selected the threshold at level $h_z$ and $\nu$ is
the original model parameter. To evaluate this limit we have to expand
$J_\nu\left(\frac{z_0}{h_z} j_{\nu,k}\right)$ near $\frac{z_0}{h_z} = 1$:
\begin{eqnarray}
J_\nu\left(\frac{z_0}{h_z} j_{\nu,k}\right) &\approx& J_\nu( j_{\nu,k}) - \left(1-\frac{z_0}{h_z}\right) \left[\nu J_\nu( j_{\nu,k}) - j_{\nu,k} J_{1+\nu}(j_{\nu,k}) \right] = (1-\frac{z_0}{h_z}) j_{\nu,k} J_{1+\nu}(j_{\nu,k}). 
\end{eqnarray}
By using this expansion we can rewrite (\ref{eq:durationPrecise}) as:
\begin{equation}
\label{eq:tdistr}
p_{h_z}^{(\nu)}(t) \approx C_1 \sum_{k=1}^{\infty} j_{\nu,k}^2
\exp\left(- \frac{j_{\nu,k}^2}{2 h_z^2} t\right) ,
\end{equation}
here $C_1$ is a normalization constant. By taking a note that $j_{\nu,k}$
are almost equally spaced, we can replace the sum by integration
\begin{eqnarray}
& p_{h_z}^{(\nu)}(t) \approx C_2 \int_{j_{\nu,1}}^{\infty} x^2 \exp\left(-\frac{x^2 t}{2 h_z^2} \right)
\mathrm{d} x = C_2 \left[\frac{h_z^2 j_{\nu,1} \exp\left(-\frac{j_{\nu,1}^2 t}{2 h_z^2}\right)}{t}+ \sqrt{\frac{\pi}{2}} \frac{h_z^3
\erfc\left(\frac{j_{\nu,1} \sqrt{t}}{\sqrt{2} h_z}\right)}{t^{3/2}} \right] .
\label{eq:durationApprox}
\end{eqnarray}
From the expression above follows that the probability density of the burst
durations in the time series obtained by solving SDE (\ref{eq:sdef}) can be
approximated by a power law with exponential cut-off. Or mathematically
\begin{eqnarray}
& p_{h_z}^{(\nu)}(t) \sim t^{-3/2} , \quad \textrm{for}\quad t \ll \frac{2 h_z^2}{j_{\nu,1}^2},\\
& p_{h_z}^{(\nu)}(t) \sim \frac{\exp\left(-\frac{j_{\nu,1}^2 t}{2 h_z^2}\right)}{t} , \quad
\textrm{for}\quad  t \gg \frac{2 h_z^2}{j_{\nu,1}^2}
\end{eqnarray}
This result is in agreement with a general property of one dimensional diffusion processes
presented in \cite{Redner2001Cambridge}, namely that the asymptotic behavior of first hitting times
is a power law $t^{-3/2}$ irrespectively of the nature of stochastic
one dimensional process or the actual mathematical expressions of the Langevin or the Fokker-Plank equations.
The exponential cutoff for longer burst durations can be explained by the direction preference of the Bessel
processes (note the positive drift term in case of $N>1$, or alternatively $\nu>-0.5$).
The actual empirical data, as shown in Fig. \ref{fig:burstDuration} (b), also has
the predicted asymptotic behavior, though the inconsistence in fitting is
clearly higher than for the model's probability density Fig. \ref{fig:burstDuration} (a).

Our empirical data set includes all trades made on NYSE, which were made from
January, 2005 to March, 2007 and involved 24 different stocks, ABT, ADM, BMY,
C, CVX, DOW, FNM, GE, GM, HD, IBM, JNJ, JPM, KO, LLY, MMM, MO, MOT, MRK,
SLE, PFE, T, WMT, XOM. We have used one hour window
moving average filter on empirical one minute return series. As we consider the model
to be universal,
i.e., applicable towards the modeling of varying financial markets and stocks, we
can consider each stocks' time series as a separate realization of the same stochastic
process. Time series are first normalized and later averaged over the whole
set. We back this approach by recalling that in \cite{Gontis2010Sciyo} we have
shown that the more sophisticated versions of (\ref{eq:return2}) may be well used to
model absolute return of different stocks from NYSE and Vilnius
Stock Exchange.

There are numerous reasons for the observed inconsistence in fitting of empirical data Fig. \ref{fig:burstDuration} (b).
Firstly, we were unable to
remove intra-day pattern from the time series. But the main reason is
that the simple stochastic model, driven by (\ref{eq:sdef}), is unable
to reproduce the full complexity of empirically observed spectral density.
In order to reproduce the correct, fractured, shape of the spectral
density one must use double stochastic model, driven by a more sophisticated
version SDE (\ref{eq:return2}), \cite{Gontis2010Sciyo}. Nevertheless,
derived equations for the burst duration distribution (\ref{eq:tdistr}) and
(\ref{eq:durationApprox}) of the general process (\ref{eq:sdef}) are in agreement
with empirical time series of return. This provides one more argument for
the further development of stochastic models based on herding behavior
of agents and nonlinear SDE (\ref{eq:sdef}).

\section{Conclusions and future work}
\label{sec:conclusions}

Reasoning of stochastic models of complex systems by the
microscopic interactions of agents is still a challenge for
researchers. Only very general models such as Kirman's herding
model in ant colony  or Bass diffusion model for new product
adoption have well established agent-based versions and can be
described by stochastic or ordinary differential equations. There
are many different attempts  of microscopic modeling in more
sophisticated systems, such as financial markets or other social
systems,  intended to reproduce the same empirically defined
properties. The ambiguity of microscopic description in complex
systems is an objective obstacle for quantitative modeling. Simple
enough agent-based models with established or expected
corresponding macroscopic description are indispensable in
modeling of more sophisticated systems. In this contribution we
discussed  various extensions and applications of Kirman's herding
model.

First of all, we modify Kirman's model introducing interevent time
$\tau(y)$ or trading activity $1/\tau(y)$ as functions of driving
return $y$. This produces the feedback from macroscopic variables
on the rate of microscopic processes and strong nonlinearity in
stochastic differential equations responsible for the long range
power-law statics of financial variables. We do expect further
development of this approach introducing the mood of chartists as
independent agent-based process.

Nonlinear SDEs derived from the agent herding model generate multifractal time series.
This gives more confidence in the modeling of multifractal series
observed in financial markets.
We derive PDF of burst duration for the basic form of nonlinear SDE (\ref{eq:sdef}).
This is in agreement with empirical time series of return.
Further investigation of burst statistics in financial markets in
comparison with analytical results from nonlinear SDE is ongoing.
This would serve as an independent method to adjust model parameters to the empirical data.

One more outcome of Kirman's herding behavior of agents is one
direction process - Bass diffusion. This simple example of
correspondence between very well established microscopic and
macroscopic modeling becomes valuable for further description of
diffusion in social systems. Models presented on the interactive
web site \cite{RizikosFizika} have to facilitate further extensive
use of computer modeling in economics, business and education.


\section*{Acknowledgment}

Work presented in this paper is supported by EU SF Project
``Science for Business and Society'', project number:
VP2-1.4-\={U}M-03-K-01-019.

We also express deep gratitude to Lithuanian Business Support Agency.

\bibliographystyle{IEEEtran}
\bibliography{Kononovicius_IARIA}

\end{document}